\documentclass[pdfLaTeX,letter]{aa}
\usepackage[varg]{txfonts}

\begin{document}

\title{Thermal conductivity of porous aggregates
}

\author{
Sota Arakawa \inst{\ref{inst1}}
\and
Hidekazu Tanaka \inst{\ref{inst2}}
\and
Akimasa Kataoka \inst{\ref{inst3}}
\and
Taishi Nakamoto \inst{\ref{inst1}}
}

\institute{
Department of Earth and Planetary Sciences, Tokyo Institute of Technology, Meguro, Tokyo, 152-8551, Japan\\
\email{arakawa.s.ac@m.titech.ac.jp} \label{inst1}
\and
Astronomical Institute, Tohoku University, Aoba, Sendai, 980-8578, Japan \label{inst2}
\and
Division of Theoretical Astronomy, National Astronomical Observatory of Japan, Mitaka, Tokyo, 181-8588, Japan \label{inst3}
}

%%\date{Received / Accepted}

\abstract
{The thermal conductivity of highly porous dust aggregates is a key parameter for many subjects in planetary science; however, it is not yet fully understood.}
{In this study, we investigate the thermal conductivity of fluffy dust aggregates with filling factors of less than $10^{-1}$.}
{We determine the temperature structure and heat flux of the porous dust aggregates calculated by $N$-body simulations of static compression in the periodic boundary condition.}
{We derive an empirical formula for the thermal conductivity through the solid network $k_{\rm sol}$ as a function of the filling factor of dust aggregates $\phi$.
The results reveal that $k_{\rm sol}$ is approximately proportional to $\phi^{2}$, and the thermal conductivity through the solid network is significantly lower than previously assumed.
In light of these findings, we must reconsider the thermal histories of small planetary bodies.}
{}


\maketitle

\section{Introduction}

Understanding the physical parameters of dust aggregates is important in planetary science.
Specifically, the thermal conductivity of dust aggregates is key for determining the thermal evolution of planetary bodies, influencing the thermal evolution pathways of both rocky and icy planetesimals \citep[e.g.,][]{Henke+2012,Sirono2017}.
The thermal evolution and activity of cometary nuclei also depend on the thermal conductivity of icy aggregates \citep[e.g.,][]{Haruyama+1993,Guilbert-Lepoutre+2011}.

Dust aggregate thermal conductivity depends on many parameters, and many previous experimental studies have researched the thermal conductivity of dust aggregates with filling factors above $10^{-1}$.
The thermal conductivity of porous aggregates in vacuum is given by two terms: the thermal conductivity through the solid network $k_{\rm sol}$ and the thermal conductivity owing to radiative transfer $k_{\rm rad}$.
\citet{Krause+2011} showed that the thermal conductivity through the solid network $k_{\rm sol}$ is exponentially dependent on the filling factor of dust aggregates $\phi$ for $0.15 < \phi < 0.54$, and concluded that the coordination number of monomer grains $C$ influences the efficiency of heat flux within the aggregates.
\citet{Sakatani+2016} revealed that $k_{\rm sol}$ is also dependent on the contact radius between monomers $r_{\rm c}$.
The thermal conductivity owing to radiative transfer $k_{\rm rad}$ is affected by the temperature of dust aggregates $T$ and the mean free path of photons $l_{\rm p}$ \citep[e.g.,][]{Schotte1960,Merrill1969}.
Moreover, $l_{\rm p}$ depends on $R$ and $\phi$ when we apply the geometrical optics approximation for the evaluation of $l_{\rm p}$ \citep[e.g.,][]{Skorov+2011,Gundlach+2012}.

There are also several theoretical studies on the thermal conductivity of dust aggregates \citep[e.g.,][]{Chan+1973,Sirono2014,Sakatani+2017}.
However, no previous research has been conducted on the thermal conductivity of porous aggregates with filling factors of less than $10^{-1}$, although \citet{Kataoka+2013b} and \citet{Arakawa+2016b} revealed that the collisional growth of dust aggregates leads to planetesimal formation via highly porous aggregates with filling factors of much less than $10^{-1}$.
Therefore, the purpose of this study is to investigate the thermal conductivity and thermal evolution of fluffy dust aggregates in protoplanetary disks.

In this letter, we calculate thermal conductivity through the solid network $k_{\rm sol}$ for highly porous aggregates with filling factors in the range of $10^{-2}$ to $10^{-1}$.
We use the snapshot data of \citet{Kataoka+2013a} for calculation of $k_{\rm sol}$.
We then validate our results through a comparison with the experimental data of \citet{Krause+2011}.
We also derive the thermal conductivity owing to radiative transfer $k_{\rm rad}$ for porous aggregates of submicron-sized monomers.
Our results show that the thermal conductivity of highly porous aggregates is significantly lower than previously assumed.

\section{Method}

\subsection{Arrangement of monomer grains}

The arrangement of monomer grains depends on the coagulation history of the aggregates.
During initial dust aggregate coagulation in protoplanetary disks, both experimental \citep[e.g.,][]{Wurm+1998} and theoretical \citep[e.g.,][]{Kempf+1999} studies have shown that hit-and-stick collisions lead to the formation of fractal aggregates with a fractal dimension $D \sim 2$, which is called ballistic cluster-cluster aggregation \citep[BCCA;][]{Meakin1991}.
Furthermore, \citet{Kataoka+2013a} performed three-dimensional numerical simulations of static compression of BCCA aggregates constituted from 16384 spherical grains using a periodic boundary condition.
In this study, we use snapshots of the compressed BCCA aggregates calculated by \citet{Kataoka+2013a}.

\subsection{Temperature structure of the dust aggregate}

To calculate the thermal conductivity through the solid network of an aggregate $k_{\rm sol}$, we have to determine the temperature of each grain in a cubic periodic boundary.
We calculate the temperature of each grain using the method of \citet{Sirono2014}.
Here, we consider one-dimensional heat flow from the lower boundary plane to the upper boundary plane.
There are three choices regarding the pair of lower and upper planes, and we calculate $k_{\rm sol}$ from three directions.
Then, we average these values for each snapshot.

\begin{figure}[h]
\centering
\includegraphics[width=0.9\columnwidth]{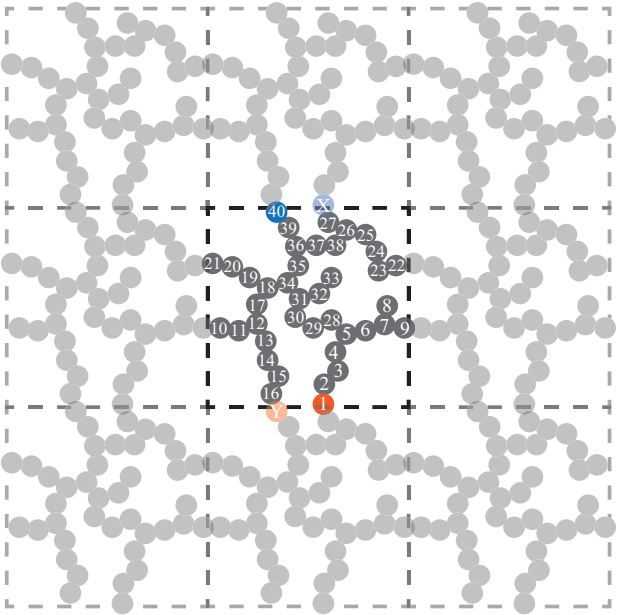}
\caption{
Sketch of a dust aggregate in a cubic periodic boundary.
The temperature of grains located on the lower (\#1 and \#Y) and upper (\#40 and \#X) boundary is set to $T_{0} + {\Delta T}/2$ and $T_{0} - {\Delta T}/2$, respectively.
The temperature of each grain is calculated by solving Eq.\ (\ref{eq1}) simultaneously for each grain.
}
\label{fig1}
\end{figure}

We define $R$ as the monomer radius and $L^{3}$ as the volume of each cubic space.
The location of the $i$-th grain $(x_{i}, y_{i}, z_{i})$ satisfies $|x_{i}| < L/2$, $|y_{i}| < L/2$, and $|z_{i}| < L/2$ for $i = 1, 2, ..., N$, where $N = 16384$ is the number of grains in the periodic boundary.
A sketch of a dust aggregate in a cubic periodic boundary is shown in Fig.\ \ref{fig1}.
Here, we assume that heat flow occurs along the z-direction.
The grains located in $- L/2 < z_{i} < - (L/2 - R)$ are on the lower boundary (\#1 in Fig.\ \ref{fig1}), and the grains located in $+ (L/2 - R) < z_{i} < + L/2$ are on the upper boundary (\#40 in Fig.\ \ref{fig1}).
If the $i$-th grain is located on the lower (upper) boundary, we add a new grain on the upper (lower) boundary.
The location of the new grain is $(x_{i}, y_{i}, z_{i} + L)$ if the $i$-th grain is located on the lower boundary (\#X in Fig.\ \ref{fig1}) and $(x_{i}, y_{i}, z_{i} - L)$ if the $i$-th grain is located on the upper boundary (\#Y in Fig.\ \ref{fig1}).
We set the temperature of grains located on the lower (\#1 and \#Y in Fig \ref{fig1}) and upper (\#40 and \#X in Fig \ref{fig1}) boundary as $T_{0} + {\Delta T}/2$ and $T_{0} - {\Delta T}/2$, respectively.

Heat flows through the monomer-monomer contacts, and for steady state conditions, the equation of heat balance at the $i$-th grain is given by
\begin{equation}
\sum_{j} F_{i, j} = 0,
\label{eq1}
\end{equation}
where $F_{i, j}$ is the heat flow from the $j$-th grain to the $i$-th grain, given by
\begin{equation}
F_{i, j} = H_{\rm c} {(T_{j} - T_{i})},
\end{equation}
where $H_{\rm c}$ is the heat conductance at the contact of two grains, and $T_{i}$ and $T_{j}$ are the temperatures of the $i$-th and $j$-th grains, respectively.
We consider the contacts not only inside the periodic boundary but also on the side boundaries (e.g., the contacts between \#9 and \#10 and between \#21 and \#22 in Fig.\ \ref{fig1}).
The heat conductance at the contact of two grains $H_{\rm c}$ is \citep{Cooper+1969}
\begin{equation}
H_{\rm c} = 2 k_{\rm mat} r_{\rm c},
\end{equation}
where $k_{\rm mat}$ is the material thermal conductivity and $r_{\rm c}$ is the contact radius of monomer grains.
The contact radius $r_{\rm c}$ depends on the monomer radius $R$ and the material parameters \citep{Johnson+1971}.
The heat conductance within a grain $H_{\rm g}$ is also given by \citep{Sakatani+2017}
\begin{equation}
H_{\rm g} = {\left( \frac{4 \pi}{3} \right)}^{1/3} k_{\rm mat} R.
\end{equation}
However, we neglect the effect of $H_{\rm g}$ because $H_{\rm g}$ is sufficiently larger than $H_{\rm c}$ for (sub)micron-sized grains.
Therefore, the temperature structure of the aggregate in the cubic periodic boundary can be calculated by solving Eq.\ (\ref{eq1}) simultaneously for all $N$ grains except lower and upper boundary grains.

\subsection{Thermal conductivity through the solid network}

Once the temperature structure is obtained, we can calculate the total heat flow at the upper boundary $\sum_{\rm upper} F_{i, j}$, where we take the sum of contacts between the upper boundary $i$-th grain and internal $j$-th grain (for the case of Fig.\ \ref{fig1}, $\sum_{\rm upper} F_{i, j} = F_{X, 27} + F_{40, 39}$).
The total heat flow at the upper boundary $\sum_{\rm upper} F_{i, j}$ can be rewritten using the thermal conductivity through the solid network $k_{\rm sol}$ as
\begin{equation}
\sum_{\rm upper} F_{i, j} = k_{\rm sol} \frac{\Delta T}{L} L^{2}.
\end{equation}
In this study, we discuss $k_{\rm sol}$ as a function of the filling factor $\phi$, and rewrite $L$ using $\phi$ as
\begin{equation}
L = {\left( \frac{4 \pi N}{3 \phi} \right)}^{1/3} R.
\end{equation}
Therefore, we obtain $k_{\rm sol}$ as a function of $\phi$ as follows:
\begin{eqnarray}
k_{\rm sol} &=& \frac{1}{L {\Delta T}} {\sum_{\rm upper} F_{i, j}}, \nonumber \\
&=& 2 k_{\rm mat} \frac{r_{\rm c}}{R} \cdot {\left( \frac{3 \phi}{4 \pi N} \right)}^{1/3} {\sum_{\rm upper} \frac{T_{j} - T_{i}}{{\Delta T}}}, \nonumber \\
&\equiv& 2 k_{\rm mat} \frac{r_{\rm c}}{R} \cdot f {(\phi)},
\label{eq7}
\end{eqnarray}
where $f {(\phi)}$ is a dimensionless function of $\phi$.
Note that the total heat flow at the lower boundary $- \sum_{\rm lower} F_{i, j}$ is clearly equal to $\sum_{\rm upper} F_{i, j}$ considering the heat balance.

\section{Results}

Here, we present the dimensionless function $f {(\phi)}$ for nine snapshots from three runs and three densities (Table \ref{table1}).
We calculated $f {(\phi)}$ in three directions for each snapshot and took the arithmetic mean values.
Note that the compressed BCCA aggregates might be isotropic if the number of monomer grains is sufficiently large; thus, we only discuss the mean values of $f {(\phi)}$.

\begin{table}[h]
\caption{Numerical calculation results.}              % title of Table
\label{table1}      % is used to refer this table in the text
\centering                                      % used for centering table
\begin{tabular}{c c c}          % centered columns (4 columns)
\hline\hline                        % inserts double horizontal lines
run & $\phi$ & $f (\phi)$ \\    % table heading
\hline\hline   % slow
    $\Box$ & $1.01 \times 10^{-2}$ & $1.44 \times 10^{-4}$ \\      % inserting body of the table
    $\Box$ & $3.02 \times 10^{-2}$ & $1.12 \times 10^{-3}$ \\
    $\Box$ & $9.91 \times 10^{-2}$ & $1.04 \times 10^{-2}$ \\
\hline   % intermediate
    $\triangle$ & $1.00 \times 10^{-2}$ & $3.86 \times 10^{-5}$ \\      % inserting body of the table
    $\triangle$ & $3.00 \times 10^{-2}$ & $5.24 \times 10^{-4}$ \\
    $\triangle$ & $1.00 \times 10^{-1}$ & $6.24 \times 10^{-3}$ \\
\hline   % fast
    \scalebox{0.8}{$\bigcirc$} & $9.98 \times 10^{-3}$ & $4.14 \times 10^{-5}$ \\      % inserting body of the table
    \scalebox{0.8}{$\bigcirc$} & $2.99 \times 10^{-2}$ & $5.15 \times 10^{-4}$ \\
    \scalebox{0.8}{$\bigcirc$} & $9.99 \times 10^{-2}$ & $9.05 \times 10^{-3}$ \\
\hline                                             %inserts single line
\end{tabular}
\end{table}

\begin{figure}[h]
\centering
\includegraphics[width=0.9\columnwidth]{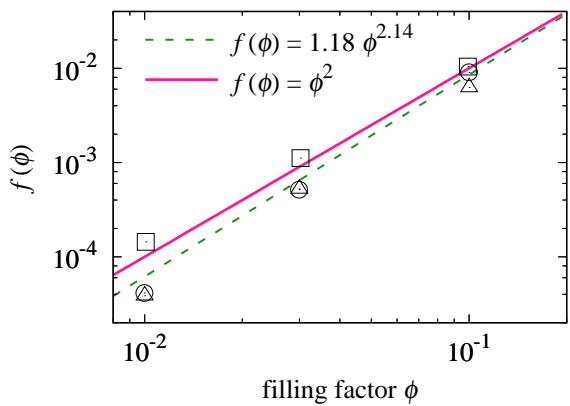}
\caption{
Fitting of the dimensionless function of thermal conductivity $f {(\phi)}$ as a function of the filling factor $\phi$.
The green dashed line is the best-fit line and the magenta solid line represents the simple function $f {(\phi)} = \phi^{2}$.
}
\label{fig2}
\end{figure}

Figure \ref{fig2} shows the dimensionless function $f {(\phi)}$ as a function of the filling factor $\phi$.
The best-fit line given by the least-squares method (green dashed line) is
\begin{equation}
f {(\phi)} = 1.18 \phi^{2.14}.
\end{equation}
Hereafter, we utilize the following more simple relationship between $f {(\phi)}$ and $\phi$ (magenta solid line)
\begin{equation}
f {(\phi)} = \phi^{2}.
\label{eq9}
\end{equation}
\citet{Sakatani+2017} predicted that $f {(\phi)} = {(2 / \pi^{2})} C \phi$, where $C$ is the coordination number.
For highly porous aggregates, the coordination number $C$ is approximately two, and the filling factor dependence on $C$ is weak.
Hence, $f {(\phi)}$ would be proportional to $\phi$ in the model of \citet{Sakatani+2017}; however, in reality, $f {(\phi)}$ is approximately proportional to $\phi^{2}$.

In the context of the thermal conductivity of colloidal nanofluid and nanocomposites, \citet{Evans+2008} revealed that thermal conductivity is strongly affected by the fraction of linear chains that contribute to heat flow in the aggregates.
The contributing grains are called backbone grains, and non-contributing grains are called dead-end grains \citep[\#8, \#32, and \#33 in Fig.\ \ref{fig1};][]{Shih+1990}.
We will discuss the effects of different fractions of backbone and dead-end grains on the thermal conductivity in future research.

\begin{figure}[h]
\centering
\includegraphics[width=0.9\columnwidth]{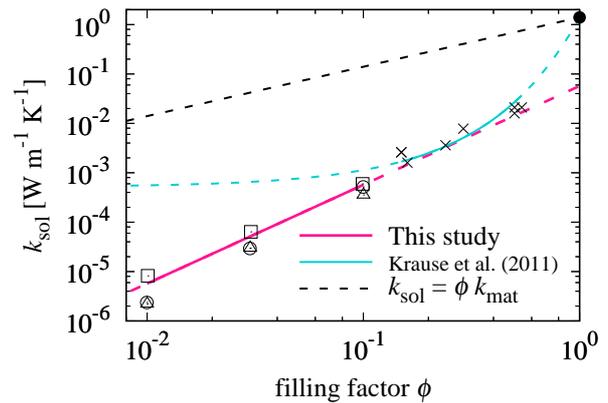}
\caption{
Experimental (crosses) and numerical (squares, triangles, and circles) data of thermal conductivity through the solid network $k_{\rm sol}$.
Our model (magenta line) was compared to the experimental fitting data of \citet[][blue curve]{Krause+2011}.
}
\label{fig3}
\end{figure}

By comparing our model to the experimental data of \citet[][]{Krause+2011}, we can confirm the validity of our model (Fig.\ \ref{fig3}).
The magenta line in Fig.\ \ref{fig3} represents the calculated thermal conductivity from Eqs.\ (\ref{eq7}) and (\ref{eq9}), the blue curve is the exponential fitting of experimental data, $k_{\rm sol} = 1.4 e^{7.91 {(\phi - 1)}}\ {\rm W}\ {\rm m}^{-1}\ {\rm K}^{-1}$ \citep{Krause+2011}, and the black dashed line is a model commonly used to study the thermal evolution of planetary bodies, that is, $k_{\rm sol} = \phi k_{\rm mat}$ \citep[e.g.,][]{Sirono2017}.
Both experimental (crosses) and numerical (squares, triangles, and circles) data are plotted.

When we consider the dust aggregates of (sub)micron-sized monomers, the contact radius between monomers $r_{\rm c}$ is given by \citep{Johnson+1971}
\begin{equation}
r_{\rm c} = {\left( \frac{9 \pi \gamma {(1 - \nu^{2})}}{2 Y R} \right)}^{1/3} R,
\label{eq10}
\end{equation}
where $\gamma = 25\ {\rm mJ}\ {\rm m}^{-2}$, $\nu = 0.17$, and $Y = 54\ {\rm GPa}$ are the surface energy, Poisson's ratio, and Young's modulus of ${\rm Si}{\rm O}_{2}$ grains, respectively \citep{Wada+2007}.
We set $R = 0.75\ \mu{\rm m}$ and $k_{\rm mat} = 1.4\ {\rm W}\ {\rm m}^{-1}\ {\rm K}^{-1}$ to the same values as \citet{Krause+2011}.
Figure \ref{fig3} clearly shows that our empirical model is applicable to the $k_{\rm sol}$ of porous aggregates with filling factors of $\phi \sim 0.1$.
Moreover, our model is applicable not only for $\phi \lesssim 0.1$ but also for the range $0.1 \lesssim \phi \lesssim 0.5$.

Note that most of the experimental data of $k_{\rm sol}$ are fitted using exponential functions of $\phi$, which pass the material thermal conductivities \citep[e.g.,][]{Krause+2011,Henke+2013}.
However, when we consider the thermal conductivity through the solid network, i.e., thermal conductivity limited by the necks between two monomers, $k_{\rm sol}$ must be given by $k_{\rm sol} \sim {(r_{\rm c} / R)} k_{\rm mat}$, even for dense dust aggregates whose filling factors are close to unity \citep[e.g.,][]{Chan+1973}.

\section{Discussion}

Finally, we evaluate the total thermal conductivity of porous icy aggregates under vacuum conditions.
Thermal conductivity through the solid network $k_{\rm sol}$ is given by
\begin{equation}
k_{\rm sol} = 2 k_{\rm mat} \frac{r_{\rm c}}{R} \phi^{2},
\end{equation}
and thermal conductivity owing to radiative transfer $k_{\rm rad}$ is given by \citep{Merrill1969}
\begin{equation}
k_{\rm rad} = \frac{16}{3} \sigma_{\rm SB} T^{3} l_{\rm p},
\end{equation}
where $\sigma_{\rm SB} = 5.67 \times 10^{-8}\ {\rm W}\ {\rm m}^{-2}\ {\rm K}^{-4}$ is the Stefan-Boltzmann constant.
We calculate the mean free path of photons in fluffy aggregates of submicron-sized grains $l_{\rm p}$ as follows:
\begin{equation}
l_{\rm p} = \frac{1}{{(\kappa_{\rm abs} + \kappa_{\rm sca})} \rho_{\rm mat} \phi},
\end{equation}
where $\kappa_{\rm abs}$ and $\kappa_{\rm sca}$ are the absorption and scattering mass opacities of monomers, respectively, and $\rho_{\rm mat} = 1.68 \times 10^{3}\ {\rm kg}\ {\rm m}^{-3}$ is the material density.
Here, the composition of icy dust aggregates is consistent with \citet{Pollack+1994}.
The total mass opacity of submicron-sized monomers $\kappa_{\rm abs} + \kappa_{\rm sca}$ is hardly dependent on the wavelength of the thermal radiation $\lambda = 2.9 \times 10^{-3}\ {(T / {\rm K})}^{-1}\ {\rm m}$ for $10^{-6}\ {\rm m} < \lambda < 10^{-4}\ {\rm m}$, and $\kappa_{\rm abs} + \kappa_{\rm sca}$ is on the order of $10^{2}\ {\rm m}^{2}\ {\rm kg}^{-1}$ \citep[e.g.,][]{Kataoka+2014}.
Then we set $l_{\rm p} = 10^{-5} \phi^{-1}\ {\rm m}$ in this study.
Note that, for the case of fluffy aggregates of submicron-sized monomers, the wavelength $\lambda$ is larger than the monomer radius $R$, even if the temperature is on the order of $10^{3}\ {\rm K}$.
Hence, we cannot apply the geometrical optics approximation for the evaluation of $l_{\rm p}$.

\begin{figure}[h]
\centering
\includegraphics[width=0.9\columnwidth]{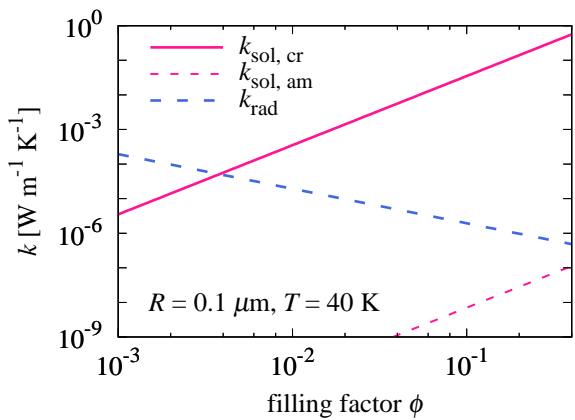}
\caption{
Comparison between $k_{\rm sol}$ (magenta solid line for crystalline icy aggregates and magenta dashed line for amorphous) and $k_{\rm rad}$ (blue dashed line).
The monomer radius is $R = 0.1\ \mu{\rm m}$ and the temperature is $T = 40\ {\rm K}$.
}
\label{fig4}
\end{figure}

Figure \ref{fig4} shows the $k_{\rm sol}$ of crystalline and amorphous icy aggregates, $k_{\rm sol, cr}$ and $k_{\rm sol, am}$, and the thermal conductivity owing to radiative transfer $k_{\rm rad}$ for aggregates composed of icy monomers with a radius of $R = 0.1\ \mu{\rm m}$ and temperature of $T = 40\ {\rm K}$.
We set $\gamma = 100\ {\rm mJ}\ {\rm m}^{-2}$, $\nu = 0.25$, and $Y = 7\ {\rm GPa}$ for icy grains \citep{Wada+2007}.
The material thermal conductivities of crystalline and amorphous grains, $k_{\rm mat, cr}$ and $k_{\rm mat, am}$, are given by $k_{\rm mat, cr} = 5.67 \times 10^{2}\ {(T / {\rm K})}^{-1}\ {\rm W}\ {\rm m}^{-1}\ {\rm K}^{-1}$ and $k_{\rm mat, am} = 7.1 \times 10^{-8}\ {(T / {\rm K})}\ {\rm W}\ {\rm m}^{-1}\ {\rm K}^{-1}$, respectively \citep{Haruyama+1993}.

For the case of $\phi < 4 \times 10^{-3}$, the thermal conductivity owing to radiative transfer $k_{\rm rad}$ is larger than the thermal conductivity through the solid network of crystalline icy aggregates $k_{\rm sol, cr}$, even if the temperature is sufficiently low ($T = 40\ {\rm K}$).
If the total thermal conductivity $k_{\rm sol} + k_{\rm rad}$ does not change substantially when crystallization occurs, then the internal temperature of icy planetesimals could still increase after crystallization, which might cause runaway crystallization due to latent heat.
In addition, when the temperature of an icy planetesimal increases, sintering can proceed inside the icy aggregate before monomer grains evaporate or melt \citep[e.g.,][]{Sirono2017}.
We cannot evaluate the contact radius $r_{\rm c}$ from Eq.\ (\ref{eq10}) when aggregates are sintered, and $k_{\rm sol}$ increases linearly as a consequence of the increase of $r_{\rm c}$.
Sintering might also affect the mechanical strength of aggregates \citep[e.g.,][]{Omura+2017} and the critical velocity for collisional growth \citep[e.g.,][]{Sirono+2017}.
Therefore, the growth pathways of icy planetesimals might be altered by sintering of icy aggregates.
Not only icy planetesimals but also rocky aggregates could experience sintering before growing into dm-sized bodies, which might explain the retainment of chondrules inside fluffy aggregates \citep{Arakawa2017}.
Note also that the total thermal conductivity might be controlled by the thermal conductivity due to gas diffusion for the case of fluffy aggregates in high gas density environments \citep[e.g., the innermost region of protoplanetary disks and/or planetary surfaces;][]{Piqueux+2009}.

In conclusion, we have revealed the filling factor dependence of the thermal conductivity of porous aggregates.
We showed that the thermal conductivity of highly porous aggregates is significantly lower than previously assumed.
In future work, we will reexamine the growth pathways of planetesimals in protoplanetary disks, and combine this with a density and thermal evolution analysis.

\begin{acknowledgements}
We thank the referee Gerhard Wurm for constructive comments.
This work is supported by JSPS KAKENHI Grant (15K05266).
S.A.\ is supported by the Grant-in-Aid for JSPS Research Fellow (17J06861).
\end{acknowledgements}

\bibliographystyle{aa}

\begin{thebibliography}{50}
\expandafter\ifx\csname natexlab\endcsname\relax\def\natexlab#1{#1}\fi

\bibitem[{{Arakawa}(2017)}]{Arakawa2017}
{Arakawa}, S. 2017, \apj, 846, 118

\bibitem[{{Arakawa} \& {Nakamoto}(2016)}]{Arakawa+2016b}
{Arakawa}, S. \& {Nakamoto}, T. 2016, \apjl, 832, L19

\bibitem[{{Chan} \& {Tien}(1973)}]{Chan+1973}
{Chan}, C.~K. \& {Tien}, C.~L. 1973, Journal of Heat Transfer, 95, 302

\bibitem[{{Cooper} {et~al.}(1969){Cooper}, {Mikic}, \& {Yovanovich}}]{Cooper+1969}
{Cooper}, M.~G., {Mikic}, B.~B., \& {Yovanovich}, M.~M. 1969, International Journal of Heat and Mass Transfer, 12, 279

\bibitem[{{Evans} {et~al.}(2008){Evans}, {Prasher}, {Fish}, {Meakin}, {Phelan}, \& {Keblinski}}]{Evans+2008}
{Evans}, W., {Prasher}, R., {Fish}, J., {et~al.} 2008, International Journal of Heat and Mass Transfer, 51, 1431

\bibitem[{{Guilbert-Lepoutre} \& {Jewitt}(2011)}]{Guilbert-Lepoutre+2011}
{Guilbert-Lepoutre}, A. \& {Jewitt}, D. 2011, \apj, 743, 31

\bibitem[{{Gundlach} \& {Blum}(2012)}]{Gundlach+2012}
{Gundlach}, B. \& {Blum}, J. 2012, \icarus, 219, 618

\bibitem[{{Haruyama} {et~al.}(1993){Haruyama}, {Yamamoto}, {Mizutani}, \& {Greenberg}}]{Haruyama+1993}
{Haruyama}, J., {Yamamoto}, T., {Mizutani}, H., \& {Greenberg}, J.~M. 1993, \jgr, 98, 15

\bibitem[{{Henke} {et~al.}(2013){Henke}, {Gail}, {Trieloff}, \& {Schwarz}}]{Henke+2013}
{Henke}, S., {Gail}, H.-P., {Trieloff}, M., \& {Schwarz}, W.~H. 2013, \icarus, 226, 212

\bibitem[{{Henke} {et~al.}(2012){Henke}, {Gail}, {Trieloff}, {Schwarz}, \& {Kleine}}]{Henke+2012}
{Henke}, S., {Gail}, H.-P., {Trieloff}, M., {Schwarz}, W.~H., \& {Kleine}, T. 2012, \aap, 537, A45

\bibitem[{{Johnson} {et~al.}(1971){Johnson}, {Kendall}, \& {Roberts}}]{Johnson+1971}
{Johnson}, K.~L., {Kendall}, K., \& {Roberts}, A.~D. 1971, Proceedings of the Royal Society of London Series A, 324, 301

\bibitem[{{Kataoka} {et~al.}(2014){Kataoka}, {Okuzumi}, {Tanaka}, \& {Nomura}}]{Kataoka+2014}
{Kataoka}, A., {Okuzumi}, S., {Tanaka}, H., \& {Nomura}, H. 2014, \aap, 568, A42

\bibitem[{{Kataoka} {et~al.}(2013{\natexlab{a}}){Kataoka}, {Tanaka}, {Okuzumi}, \& {Wada}}]{Kataoka+2013a}
{Kataoka}, A., {Tanaka}, H., {Okuzumi}, S., \& {Wada}, K. 2013{\natexlab{a}}, \aap, 554, A4

\bibitem[{{Kataoka} {et~al.}(2013{\natexlab{b}}){Kataoka}, {Tanaka}, {Okuzumi}, \& {Wada}}]{Kataoka+2013b}
{Kataoka}, A., {Tanaka}, H., {Okuzumi}, S., \& {Wada}, K. 2013{\natexlab{b}}, \aap, 557, L4

\bibitem[{{Kempf} {et~al.}(1999){Kempf}, {Pfalzner}, \& {Henning}}]{Kempf+1999}
{Kempf}, S., {Pfalzner}, S., \& {Henning}, T.~K. 1999, \icarus, 141, 388

\bibitem[{{Krause} {et~al.}(2011){Krause}, {Blum}, {Skorov}, \& {Trieloff}}]{Krause+2011}
{Krause}, M., {Blum}, J., {Skorov}, Y.~V., \& {Trieloff}, M. 2011, \icarus, 214, 286

\bibitem[{{Meakin}(1991)}]{Meakin1991}
{Meakin}, P. 1991, Reviews of Geophysics, 29, 317

\bibitem[{{Merrill}(1969)}]{Merrill1969}
{Merrill}, R.~B., 1969, Nasa Technical Note, D-5063

\bibitem[{{Omura} \& {Nakamura}(2017)}]{Omura+2017}
{Omura}, T. \& {Nakamura}, A.~M. 2017, \planss, in press [\eprint[arXiv]{1708.08038}]

\bibitem[{{Piqueux} \& {Christensen}(2009)}]{Piqueux+2009}
{Piqueux}, S. \& {Christensen}, P.~R. 2009, \jgr, 114, E09005

\bibitem[{{Pollack} {et~al.}(1994){Pollack}, {Hollenbach}, {Beckwith}, {Simonelli}, {Roush}, \& {Fong}}]{Pollack+1994}
{Pollack}, J.~B., {Hollenbach}, D., {Beckwith}, S., {et~al.} 1994, \apj, 421, 615

\bibitem[{{Sakatani} {et~al.}(2017){Sakatani}, {Ogawa}, {Iijima}, {Arakawa}, {Honda}, \& {Tanaka}}]{Sakatani+2017}
{Sakatani}, N., {Ogawa}, K., {Iijima}, Y.-i., {et~al.} 2017, AIP Advances, 7, 015310

\bibitem[{{Sakatani} {et~al.}(2016){Sakatani}, {Ogawa}, {Iijima}, {Arakawa}, \& {Tanaka}}]{Sakatani+2016}
{Sakatani}, N., {Ogawa}, K., {Iijima}, Y.-i., {Arakawa}, M., \& {Tanaka}, S. 2016, \icarus, 267, 1

\bibitem[{{Schotte}(1960)}]{Schotte1960}
{Schotte}, W. 1960, AIChE Journal, 6, 63

\bibitem[{{Shih} {et~al.}(1990){Shih}, {Shih}, {Kim}, {Liu}, \& {Aksay}}]{Shih+1990}
{Shih}, W.-H., {Shih}, W.~Y., {Kim}, S.-I., {Liu}, J., \& {Aksay}, I.~A. 1990, \pra, 42, 4772

\bibitem[{{Sirono}(2014)}]{Sirono2014}
{Sirono}, S.-i. 2014, Meteoritics and Planetary Science, 49, 109

\bibitem[{{Sirono}(2017)}]{Sirono2017}
{Sirono}, S.-i. 2017, \apj, 842, 11

\bibitem[{{Sirono} \& {Ueno}(2017)}]{Sirono+2017}
{Sirono}, S.-i. \& {Ueno}, H. 2017, \apj, 841, 36

\bibitem[{{Skorov} {et~al.}(2011){Skorov}, {van Lieshout}, {Blum}, \& {Keller}}]{Skorov+2011}
{Skorov}, Y.~V., {van Lieshout}, R., {Blum}, J., \& {Keller}, H.~U. 2011, \icarus, 212, 867

\bibitem[{{Wada} {et~al.}(2007){Wada}, {Tanaka}, {Suyama}, {Kimura}, \& {Yamamoto}}]{Wada+2007}
{Wada}, K., {Tanaka}, H., {Suyama}, T., {Kimura}, H., \& {Yamamoto}, T. 2007, \apj, 661, 320

\bibitem[{{Wurm} \& {Blum}(1998)}]{Wurm+1998}
{Wurm}, G. \& {Blum}, J. 1998, \icarus, 132, 125

\end{thebibliography}

\end{document}